\documentclass{article}

\usepackage[preprint,nonatbib]{neurips_2024}
\usepackage[numbers,sort&compress]{natbib}

\usepackage[utf8]{inputenc} 
\usepackage[T1]{fontenc}    
\usepackage{hyperref}       
\usepackage{url}            
\usepackage{amssymb}        
\usepackage{booktabs}       
\usepackage{amsfonts}       
\usepackage{graphicx}       
\usepackage{nicefrac}       
\usepackage{microtype}      
\usepackage{xcolor}         
\usepackage{multirow} 
\usepackage{enumitem}
\usepackage{amsmath} 

\title{A Masked Representation Learning to Model Cardiac Functions Using Multiple Physiological Signals}

\author{%
  \textbf{Seong-A Park} $^{1}$\thanks{These authors contributed equally.}\quad
  \textbf{Jong-Eui Chae} $^{2}$\footnotemark[1]\quad
  \textbf{Sungdong Kim} $^{3}$ \quad \\[1pt]
  \textbf{Hyung-Chul Lee} $^{1,2}$ \quad
  \textbf{Hyun-Lim Yang} $^{1,2}$\thanks{Corresponding author.} \\
  \\
  \multicolumn{1}{c}{$^{1}$Seoul National University Hospital \quad $^{2}$Seoul National University \quad $^{3}$KAIST AI} \\[1pt]
  \multicolumn{1}{c}{\texttt{\{seonga, hlyang\}@snu.ac.kr}}
}

\begin{document}

\maketitle
\begin{abstract}
    In clinical settings, monitoring hemodynamics is crucial for managing patient prognosis, necessitating the integrated analysis of multiple physiological signals. 
    While recent research has analyzed single signals such as electrocardiography (ECG) or photoplethysmography (PPG), there has yet to be a proposal for an approach that encompasses the complex signal analysis required in actual clinical scenarios. 
    In this study, we introduce the SNUPHY-M (Seoul National University hospital PHYsiological signal Masked representation learning) model extracts physiological features reflecting the electrical, pressure, and fluid characteristics of the cardiac cycle in the process of restoring three masked physiological signals based on self-supervised learning (SSL): ECG, PPG, and arterial blood pressure (ABP) signals.
    By employing multiple physical characteristics, the model can extract more enriched features only using non-invasive signals. 
    We evaluated the model’s performance in clinical downstream tasks such as hypotension, stroke volume, systolic blood pressure, diastolic blood pressure, and age prediction. 
    Our results showed that the SNUPHY-M significantly outperformed supervised or SSL models, especially in prediction tasks using non-invasive signals. 
    To the best of our knowledge, SNUPHY-M is the first model to apply multi-modal SSL to cardiovascular analysis involving ECG, PPG, and ABP signals.
    This approach effectively supports clinical decision-making and enables precise diagnostics, contributing significantly to the early diagnosis and management of hemodynamics without invasiveness. 
\end{abstract}

\section{Introduction}
Hemodynamic monitoring is defined as the real-time evaluation of a patient's cardiovascular function, including parameters such as cardiac output, systemic vascular resistance, and blood pressure \cite{hemodynamicdefinition}. 
This process is conducted through the observation and measurement of a range of physical attributes inherent to physiological signals, including electrocardiography (ECG), which is used to record the heart's electrical activity; photoplethysmography (PPG), which is employed to detect changes in blood flow through near-infrared light; and arterial blood pressure (ABP), which is used to measure intra-arterial pressure directly \cite{hemodynamicmonitoringecgppgabp}. 
Continuous monitoring and comprehensive interpretation of dynamically interacting physiological signals are essential for effective early hemodynamic management, which has been shown to significantly reduce mortality \cite{hemodynamiceffect,hemodynamicml}.
This is of particular importance in the context of patient evaluation and the implementation of timely clinical interventions in high-risk environments, such as the operating room or the intensive care unit \cite{hemodynamiceffect,hemodynamicml,hemodynamicconsensus}.
For example, ECG, PPG, and ABP signals are employed to assess cardiac rhythm, alterations in blood flow, and fluctuations in blood pressure, respectively.
Changes in heart rate affect blood pressure and oxygen saturation, and fluctuations in blood pressure reflect blood flow velocity and cardiac pumping function. 
The interaction of these various physiological variables is intricate, and they undergo dynamic changes based on the patient's condition and response. 
This renders the analysis of these variables and the modeling of cardiac function highly demanding.

\begin{figure*}[t]
\centering
\includegraphics[width=\textwidth, height=6cm]{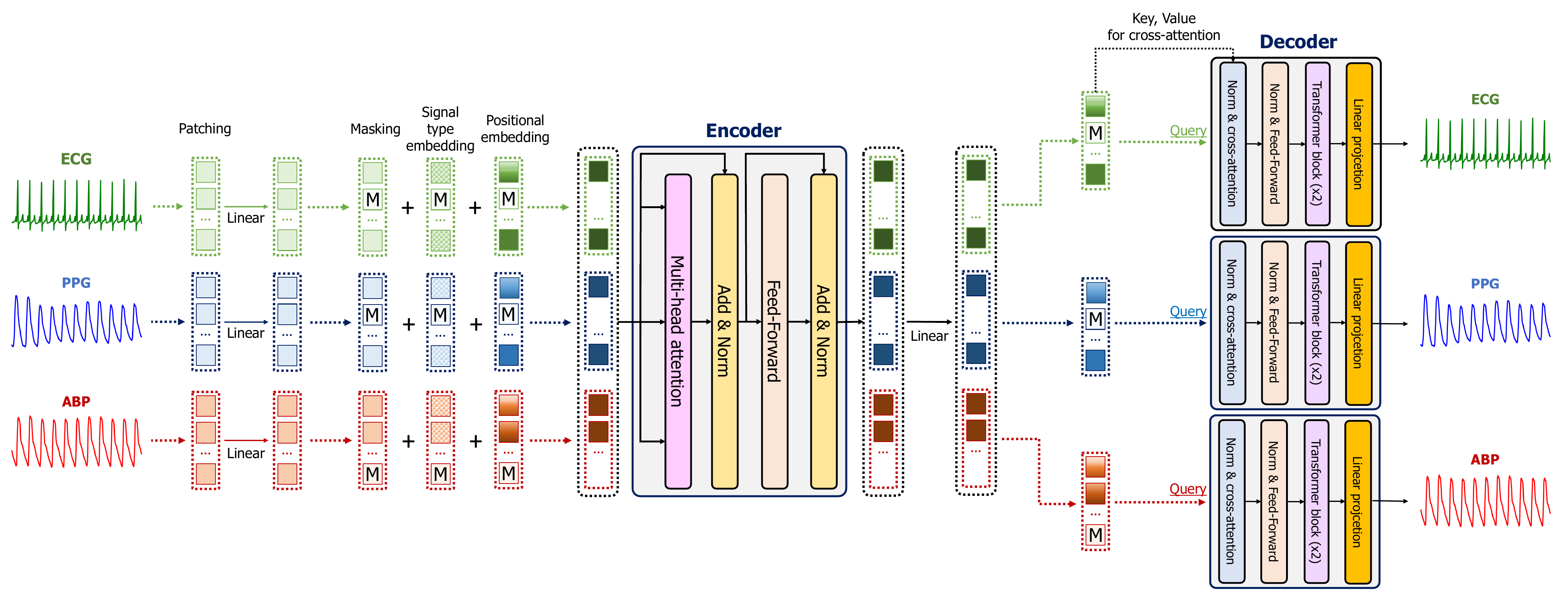} 
\caption{Overview of the SNUPHY-M. SNUPHY-M with joint encoding and individual decoding of physiological signals.}
\label{fig_overall_architecture}
\end{figure*}

In recent years, there has been a notable increase in interest regarding the utilization of deep learning (DL) models for the analysis of intricate physiological signals \cite{ecgcnnsl,attnslppg,attnslmiccai,aaaippgtoecg}.
The acquisition and labeling of clinical data are highly resource-intensive, and constructing task-specific datasets to develop models that meet various clinical requirements is a highly challenging undertaking.
To address these limitations, self-supervised learning (SSL) methodologies have emerged as a promising alternative for the generation of high-quality representation vectors for physiological signals \cite{AppleECGPPGCL,BEITECG,biot,wearableecgssl}.
SSL entails the identification of distinctive patterns or structures within data without the necessity for labels.
However, the majority of these attempts have been limited to the analysis of a single signal.
As ECG, PPG, and ABP signals are all independent indicators of a single cardiac cycle represented by heartbeats, integrating multiple physiological signals for learning may prove to be a more effective method for capturing and learning the subtle changes that may be difficult to detect with a single signal alone \cite{multiplefeaturestransfusion,multiplesignalhypotension}.
For example, the ECG signal can be used to detect functional abnormalities in the heart, such as arrhythmias, while the PPG signal provides insights into blood flow, and the ABP signal measurement allows for the indirect estimation of vascular resistance and volume, facilitating the detection of various clinical anomalies.

Furthermore, in actual clinical settings, non-invasive monitoring is typically preferred.
This has been shown to help protect patients from unnecessary infections and improve patient compliance \cite{ramsay2020non}. 
The affordability and accessibility of non-invasive monitoring devices facilitate the expansion of monitoring to a greater number of patients and environments.

To address the limitations and requirements, we introduce SNUPHY-M (Seoul National University hospital PHYsiological signal Masked representation learning), a multi-modal masked autoencoder-based SSL framework that integrates the learning of multiple physiological signals.
It enables the model to capture representations that maximize the unique characteristics and complementary information of each signal. 
SNUPHY-M employs diversified masking and reconstruction strategies considering multiple physiological signals, thereby guiding the model to comprehend the intricate interrelationships between signals. 
In order to enhance the clinical utility, each signal has been assigned its own decoder, thus enabling the generation of a representation even in the absence of some signals.
Using the proposed framework, we pre-train a model and subsequently design and evaluate various downstream tasks using data obtained from real clinical environments, thereby assessing the applicability of this approach in practical medical settings.
The contributions of this work are as follows: 
\begin{itemize}[left=0pt]
    \item We propose a representation learning framework that simultaneously considers multiple physiological signals, focusing on cardiac physiological functions.
    \item We introduce inter-, intra-, and signal-masking strategies suitable for analyzing physiological signals, where time synchronization is crucial.
    \item We used distinct decoders for each signal to allow for variable-length inputs, thereby enhancing the clinical utility of the model.
    \item We evaluate the utility of the model by validating its performance on five key clinical prediction tasks using non-invasive signals from real clinical environments and external validations.
\end{itemize}

\section{Related Work}
\subsection{Deep learning model for physiological signals} 
Recent studies have demonstrated the efficacy of utilizing physiological signals to predict cardiovascular conditions, including hypotension and stroke volume (SV).
In the context of hypotension prediction tasks, \citet{sbpecgppg} showed that systolic blood pressure (SBP) can be calculated non-invasively during surgical procedures by utilizing ECG and PPG signals and applying multiple linear regression.
Moreover, the findings of \citet{attnslppg} indicate that applying a convolutional neural network (CNN) model with an attention mechanism, utilizing solely the PPG signal can facilitate more precise blood pressure estimation.
In tasks about SV prediction, \citet{adultapco} demonstrated that reduced invasiveness can be achieved by using an attention-based CNN model applied to the ABP signal.
Moreover, \citet{ecgppgtobp} employed a CNN-long short-term memory model to predict blood pressure from the PPG signal.
These endeavors are intended to diminish reliance on invasive measurements by deriving cardiovascular metrics from less- or non-invasive signals and to expand clinical utilities.
However, these methods still necessitate substantial labeled data, which is challenging to retain in clinical settings.


\subsection{Self-supervised learning with physiological signals} 
The successful emergence of SSL without the necessity of extensive labeling data has facilitated its application to physiological signals. Studies such as \citet{BEITECG}, \citet{MAEFE}, and \citet{STMEM} have applied masked representation learning, such as BEiT\cite{beit} or MAE\cite{MAE}, to 12-lead ECG signals and achieved improvements in signal reconstruction and diagnostic performance. \citet{AppleECGPPGCL} extended SSL to wearable ECG and PPG signals using contrastive learning. On the other hand, the majority of prior works focus on single-modality signals, which can lead to a deficiency of hemodynamic information, decreased predictive performance, and increased vulnerability to noise. To overcome these limitations, an integrated embedding that effectively models the complex interrelationships among signals is required. 

\begin{figure}[t]
\centering
\includegraphics[width=\columnwidth, height=3.5cm]{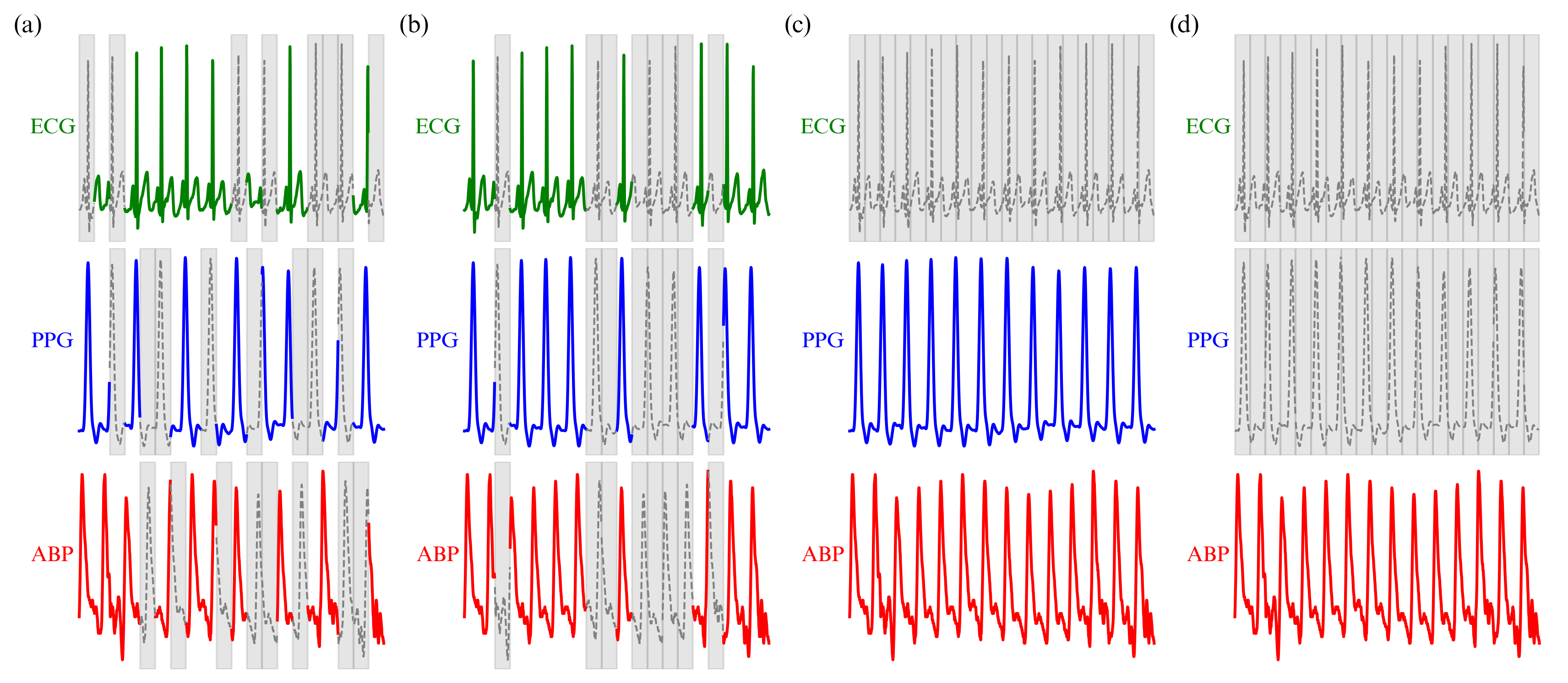} 
\caption{(a) Inter-masking and (b) Intra-masking methods, both with a $40\%$ masking ratio. (c) and (d) depict Signal-masking methods, where the entire duration of one signal (c) or two signals (d) is masked. Gray-shaded areas represent the masked regions.}
\label{fig_masking_strategy}
\end{figure}

\section{Method}
Figure \ref{fig_overall_architecture} depicts the overall architecture of the proposed framework.
The framework is structured to learn the relationships among signals that, despite their different characteristics, share a common cardiac cycle. 
This enables effective modeling of cardiovascular dynamics even in clinical settings where certain signals are missing or limited.

\subsection{Patching and projection}
The signal type is indicated by $S_i$, where $i$ denotes the index of the respective signal. 
The set of signals $S$ is composed of $\{\text{ECG}, \text{PPG}, \text{ABP}\}$.
The signal $S_i$ is segmented into discrete patches, denoted by $P_{i,j}$, where $j$ indicates a specific patch within the signal $S_i$. 
Each patch $P_{i,j}$ has a length of $p$. 
The number of patches, denoted by $J$, is calculated using the formula $J = \text{length}(S_i) / p$, 
where $J \in \mathbb{Z}$ represents the patch size as $S_i$.


The length of patches, denoted as $p$, was set to 0.5-seconds to ensure the incorporation of the distinctive attributes of each signal. 
This choice was based on experiments conducted with various lengths (0.5s, 1s, 2s, 2.5s) as detailed in Appendix B.
With $p$ of 0.5-seconds, ECG patches might encompass features such as the QRS complex, S-peaks, T-peaks, P-peaks, PPG patches can include periodic waveform characteristics corresponding to changes in blood flow, and ABP patches can capture arterial pressure pulsations on a single beat.
This configuration ensures that local cardiac contraction or relaxation features are effectively represented.
Subsequently, a linear projection is applied to each signal $P_{i,j}$ from signal $S_i$, in order to transform the data into a higher-dimensional feature space. 
This linear projection, denoted by $f_i$, maps the original data from $f_i(\cdot): \mathbb{R}^{n \times J \times p} \mapsto \mathbb{R}^{n \times J \times d}$, where $d$ is the dimension of the feature space and $n$ denotes the number of samples or batch size.

\subsection{Signal type and positional embedding}

The signal type embedding ($t_i$) is introduced to enable the encoder to encompass cardiac physiological functions by considering diverse signals such as ECG, PPG, and ABP, while still learning the distinctive characteristics of each signal.
This embedding is added to each projected signal ($f_i(S_i)$) to guarantee that the same signal type is consistently situated within a comparable distribution. 
The distribution of $t_i \in \mathbb{R}^{J \times d}$ is aligned with the standard deviation of each signal.
Subsequently, positional encoding ($\rho \in \mathbb{R}^{J \times d}$) is applied to the projected data to incorporate additional temporal context.
The sinusoidal positional encoding is employed.
Finally, the embedded vector ($Z_i$) after patching, projection, signal type embedding, and positional embedding is as follows: $Z_i = f_i(S_i)+t_i+\rho$.

\subsection{Masking strategy}

Three masking strategies were introduced to facilitate the discovery of intricate relationships between heterogeneous signals: inter-, intra-, and signal-masking.
These strategies entail the creation of a masked index, denoted by $\mathcal{M}$, which is used for further processing of the signals.

\paragraph{\textbf{Inter-masking}}
As illustrated in Figure \ref{fig_masking_strategy}-(a), distinct patches are masked across disparate signals, prompting the model to infer absent information in one signal by capitalizing from other signals. 
This prompts the model to examine the interdependencies and complementary information between the signals, thereby enhancing its capacity to discern intricate physiological characteristics.

\paragraph{\textbf{Intra-masking}}
As illustrated in Figure \ref{fig_masking_strategy}-(b), entails the synchronous masking of time-aligned patches across all signals.
This synchronization of information loss enables the model to learn the intra-dependencies within a signal and simultaneously consider multiple signals.

\paragraph{\textbf{Signal-masking}}
To enhance the model's capacity to reconstruct previously unseen data using the remaining signals, we deliberately obfuscate the entirely of a specific input signal. 
Figure \ref{fig_masking_strategy}-(c) and (d) illustrate examples of signal-masking, in which a single or two input signals are selected and entirely masked among three input signals.
This enables the effective inference and reconstruction of information within a model, even in the absence of a substantial proportion of the data.

\subsection{Encoder and decoder training}

The SNUPHY-M framework was inspired by Multi-MAE \cite{2022multimae}, and employed the ViT \cite{VIT} architecture to encode non-masked patches of physiological signals.
The non-masked signals, denoted by $Z'_{i}$, which are sequences of $J'$ non-masked patches following the $\mathcal{M}$, are concatenated and used as the input to the encoder.
The input to the encoder, $X$, is constructed as follows:
\begin{align*}
    X &= (Z'_{\text{ECG}} \parallel Z'_{\text{PPG}} \parallel Z'_{\text{ABP}}) \in \mathbb{R}^{n \times (3 \times J') \times d}
\end{align*}
where "$\parallel$" represents concatenation and $J'$ denotes the number of non-masked patches in $Z$. 


The encoder module, designated as $g$, comprises a multi-head attention layer (MHA) and a feed-forward layer. 
The MHA layer applies multiple attention heads in parallel, enabling the model to focus on various features of the input data and capture richer representations. 
The output of MHA is added to the original input through a residual connection, followed by layer normalization to enhance model stability. 
This is then passed through a feed-forward layer, where residual connections and layer normalization are again applied, ensuring stable gradient convergence during backpropagation. 


Prior to its entry into the decoder ($g^{-1}$), the output of the encoder ($g(X)$) is mapped to a lower-dimensional space through a linear projection ($h$) with the objective of reducing computational complexity and enhancing the model's generalization capabilities \cite{MAE}.
The output of linear projection, $h$, is $E = h(g(X)) \in \mathbb{R}^{3J' \times d}$.
A distinct decoder is employed for each signal, designed to reconstruct the masked patches.
Accordingly, the embedding vector $E$ is partitioned into three vectors: ($E_{\text{ECG}} = [e_1, \cdots, e_{J'}]$, $E_{\text{PPG}} = [e_{J'+1}, \cdots, e_{2J'}]$, $E_{\text{ABP}} = [e_{2J'+1}, \cdots, e_{3J'}]$)
Each encoded embedding ($E_i$) is fed to the decoder ${g_i}^{-1}$ along with masked embedding ($\mu \in \mathbb{R}^d$) generated according to $\mathcal{M}$.
Let the input of each decoder as $E'_{i} = E_i \boxplus \mu$, where $\boxplus$ represents patch-wise merge according to $\mathcal{M}$.
Subsequently, positional encoding ($\rho$) is applied again to the combined signals to preserve the spatial relationships within the data.
In order to capture the interrelated characteristics of multiple signals, multi-head cross-attention mechanisms are employed for the decoders.
For each signal, the decoder receives the signal-specific encoded embedding ($E_i$) as the query input, while the combined encoded features from all signals ($E$) are used as both keys and values in the cross-attention operation. 
This enables the decoder to selectively access and integrate pertinent information from the surrounding signals and their characteristics when reconstructing the masked patches of each specific signal, thereby ensuring that the most relevant details from other signals are utilized for reconstruction. 
The output from the cross-attention passes through a transformer block and is then mapped to the reconstruction dimension via a linear projection.
The transformer block has the same structure as the encoder ($g$).
Finally, the output produced by each decoder is denoted as $\hat{S}_i = {g_i}^{-1}(E'_{i})$, representing the reconstructed signal.


To train the encoder and decoder to accurately reconstruct the physiological signal, we present a tailored loss function that employs root mean squared error (RMSE) and Pearson correlation coefficient (PCC).
The RMSE and PCC for $S_i$ are defined as:
\[
\text{RMSE}(S_i, \hat{S}_i) = \sqrt{\frac{1}{N_i} \sum_{t=1}^{N_i} (S_i(t) - \hat{S}_i(t))^2}
\]

\[
\text{PCC}(S_i, \hat{S}_i) = \frac{\sum_{t=1}^{N_i} (S_i(t) - \bar{S}_i)(\hat{S}_i(t) - \bar{\hat{S}}_i)}{\sqrt{\sum_{t=1}^{N_i} (S_i(t) - \bar{S}_i)^2 \sum_{t=1}^{N_i} (\hat{S}_i(t) - \bar{\hat{S}}_i)^2}}
\]

The loss function is the sum of the RMSE and PCC for each signal, where the coefficients $\alpha$ and $\beta$ were determined experimentally and set to 0.8 and 0.2, respectively.

\begin{align*}
\text{Loss} = &\ \alpha \left(\text{RMSE}(S_{\text{ECG}}, \hat{S}_{\text{ECG}}) \right. \\
&\qquad \left. + \text{RMSE}(S_{\text{PPG}}, \hat{S}_{\text{PPG}}) + \text{RMSE}(S_{\text{ABP}}, \hat{S}_{\text{ABP}})\right) \\
& + \beta \left(\text{PCC}(S_{\text{ECG}}, \hat{S}_{\text{ECG}}) \right. \\
&\qquad \left. + \text{PCC}(S_{\text{PPG}}, \hat{S}_{\text{PPG}}) + \text{PCC}(S_{\text{ABP}}, \hat{S}_{\text{ABP}})\right)
\end{align*}


While RMSE focuses on minimizing the magnitude of prediction errors, PCC captures the linear correlation between predicted and reference signals, emphasizing pattern alignment. 
Incorporating both metrics into the loss function enables signal reconstruction that maintains low error while preserving the temporal and structural characteristics of the signal.

\section{Experiments}
\subsection{Datasets}
\paragraph{\textbf{Ethical approval and data collection}}
The collection of data and subsequent analysis were approved by the institutional review board of  Seoul National University Hospital (H-2307-110-1449).
7,849 surgical procedures were used for SSL training, and 1,386 were reserved for downstream evaluation. 
A detailed description of the data collection and preprocessing is presented in Appendix A.

\paragraph{\textbf{Downstream datasets}} 
Two downstream datasets were constructed: the internal dataset and the VitalDB open dataset \cite{vitaldb}.
Each dataset was recorded in different devices, an IntelliVue\textsuperscript{TM} patient monitor (Phillips, Amsterdam, The Netherlands) and a Solar8000 patient monitor (GE Healthcare, Chicago, Illinois, USA), respectively.
Notably, the PPG signals may exhibit variations in signal characteristics depending on the type of monitoring device used, which could potentially influence downstream analyses.
The SNUPHY-M was also evaluated for its blood pressure estimation performance on the MIMIC-III waveform dataset \cite{MIMIC}, to assess its versatility and effectiveness across diverse datasets collected from various monitoring systems.
Details of downstream dataset construction are presented in Appendix A.

\subsection{Experiments setup}
\paragraph{\textbf{Pre-training and supervised training setup}} 
The $40\%$ inter-, intra-, and signal-masking were employed for all signals, based on the empirical findings detailed in Appendix B.
The training employed a batch size 4,096 and a learning rate 0.001.
Twelve encoder modules were used, and the feature dimension of the multi-head attention layer and the feed-forward layer was set to 256.
Twelve modules were used for the decoder, with the feature dimension of each attention layer and transformer block set to 128.
In the SNUPHY-M, the loss is updated only after processing 10 distinct batches.
These batches incorporate various masking strategies, including intra-masking, inter-masking, and signal-masking. 
Signal-masking is applied to individual signals, such as ECG, PPG, and ABP, as well as to combinations of signals, including ECG and PPG, PPG and ABP, and ECG and ABP. 
Gradient accumulation was employed to make the model learn various masking patterns, which enhances its generalization ability under different conditions.


The single SSL and supervised learning (SL) models were employed to validate the superiority of SNUPHY-M.
For single SSL models, which are designed to process a single type of physiological signal, the architecture does not incorporate the signal embedding and cross-attention mechanisms present in the SNUPHY-M. 
During pre-training, each single SSL model learned from each signal with random masking applied at the same masking ratio, following the SNUPHY-M. 
In the case of SL, solely the encoder part of the SNUPHY-M was employed to perform predictions for all tasks. 
Unlike SNUPHY-M, in the SL and single SSL models, the loss is computed and updated on each training batch during the optimization process. 

\paragraph{\textbf{Downstream evaluation}} 
Five representative physiological signal prediction tasks were considered: one classification task for predicting hypotension, and four regression tasks for predicting stroke volume (SV), systolic blood pressure (SBP), diastolic blood pressure (DBP), and age \cite{attnslmiccai,AppleECGPPGCL}.
Note that three different prediction windows (5-, 10-, and 15-min) were considered for the hypotension prediction task.
To ensure objectivity and rigor, all evaluations were performed through external validation using publicly available datasets.
SNUPHY-M was evaluated against single SSL and SL models on non-invasive physiological signals. 
It was also evaluated by varying the input signals used for inference, thereby assessing the performance in scenarios where data is scarce, as would be the case in actual clinical settings. 
The linear probing using output from the encoder of SNUPHY-M was employed for downstream evaluation.
The detailed configurations of linear probing are provided in Appendix A.
The performance was evaluated using the area under the receiver operating characteristic curve (AU-ROC) for the classification task and mean absolute error (MAE) for other tasks.
To ascertain the statistical significance of the performance, a DeLong test was conducted on the classification task, while a t-test with bootstrapping was performed on the regression tasks.
A $p$-value of less than 0.01 was considered statistically significant.

\subsection{Results} 
\paragraph{\textbf{Linear probing}}

\begin{table*}[ht]
\caption{Linear probing performance comparison on internal dataset. AU-ROC for hypotension prediction, MAE for other tasks.}
\resizebox{\textwidth}{!}{%
\centering
\begin{tabular}{@{}cccccccccc@{}}
\toprule
\multirow{2}{*}{Model} & \multirow{2}{*}{\begin{tabular}[c]{@{}c@{}}Training \\ signals\end{tabular}} & \multirow{2}{*}{\begin{tabular}[c]{@{}c@{}}Inference \\ signals\end{tabular}} & \multicolumn{3}{c}{Hypotension} & \multirow{2}{*}{SBP (mmHg)} & \multirow{2}{*}{DBP (mmHg)} & \multirow{2}{*}{Age (years)} \\ 
 &  &  & 5-min & 10-min & 15-min &  &  &  \\
\toprule
SL & ECG & ECG & 0.505 & 0.506 & 0.478 & 10.085 (12.419) & 7.456 (9.296) & 14.046 (10.707) \\
 & PPG & PPG & 0.745 & 0.792 & 0.801 & 10.086 (12.419) & 7.532 (9.491) & 11.633 (9.064) \\
 & ECG, PPG & ECG, PPG & 0.480 & 0.466 & 0.517 & 10.090 (12.419) & 7.458 (9.295)  & 14.045 (10.746) \\  

Single SSL & ECG & ECG & 0.687 & 0.766 & 0.727 & 10.140 (12.479) &  7.352 (9.151) & 12.659 (9.084) \\
 & PPG & PPG & 0.773 & 0.772 & 0.752 & 9.926 (12.202) & 7.286 (9.145) & 11.978 (9.166) \\ 

SNUPHY-M & ALL & ECG & 0.708 & 0.812 & 0.740 & 10.167 (12.490) & 7.314 (9.081) & 12.201 (9.648) \\
 & ALL & PPG & \textbf{0.887} & 0.856 & 0.891 & \textbf{9.025 (11.388)} & 6.766 (8.712) & 10.337 (8.337) \\
 & ALL & ECG, PPG & 0.871 & \textbf{0.879} & \textbf{0.896} & 9.092 (11.462) & \textbf{6.672 (8.605)} & \textbf{10.126 (8.069)} \\ \bottomrule
\end{tabular}
}
\label{tbl_full_results_IntelliVue}
\end{table*}

Table \ref{tbl_full_results_IntelliVue} summarizes the performance of various models on the internal dataset, which was evaluated using the same monitoring device utilized for pre-training. 
The SNUPHY-M demonstrates consistently superior performance to SL models across all tasks.
Specifically, in the 5-, 10-, and 15-minute hypotension prediction, the SNUPHY-M with ECG and PPG signals achieved AU-ROC scores of 0.871, 0.879, and 0.896, whereas the SL model showed 0.480, 0.466, and 0.517, respectively (all $p<0.01$). 
When trained and inferred using ECG and PPG signals, the SNUPHY-M achieved the MAE of 9.092 mmHg for SBP estimation and 6.672 mmHg for DBP estimation, which was significantly lower than the SL model's best performance of 10.090 mmHg for SBP estimation and 8.314 mmHg for DBP estimation ($p<0.01$).
In the age prediction, the SNUPHY-M achieved the MAE of 10.126 years, which was significantly lower than the 14.045 years obtained by the SL model using all signals ($p<0.01$).
SNUPHY-M demonstrates superior performance compared to single SSL methods across all tasks except for the SBP estimation task using ECG signal.
Specifically, for the 5-minute, 10-minute, and 15-minute hypotension prediction, SNUPHY-M with ECG signal achieves higher AU-ROC of 0.708, 0.812, 0.740, than that of the single SSL model with 0.687, 0.766, and 0.727, respectively (all $p<0.01$).
Similarly, SNUPHY-M significantly outperforms the single SSL model with AU-ROC of 0.887, 0.856, and 0.891, compared to those of the single SSL model with 0.773, 0.772, and 0.752, using PPG (all $p<0.01$), highlighting its consistent and notable effectiveness.
In the SBP estimation task, the single SSL model outperformed SNUPHY-M when only ECG signals were available (MAE of 10.140 vs. 10.167 mmHg; $p<0.01$), while SNUPHY-M outperformed the single SSL model when only the PPG signal was available (MAE of 9.926 vs. 9.025 mmHg; $p<0.01$). 
In the DBP estimation task, SNUPHY-M achieved a lower MAE compared to the single SSL model with 7.314 vs. 7.352 mmHg using ECG signal and 6.766 vs. 7.284 mmHg using PPG signal (all $p<0.01$). 

\begin{table*}[ht]
\caption{Linear probing performance comparison on VitalDB dataset. AU-ROC for hypotension prediction, MAE for other tasks.}
\resizebox{\textwidth}{!}{%
\centering
\begin{tabular}{@{}ccccccccccc@{}}
\toprule
\multirow{2}{*}{Model} & \multirow{2}{*}{\begin{tabular}[c]{@{}c@{}}Training \\ signals\end{tabular}} & \multirow{2}{*}{\begin{tabular}[c]{@{}c@{}}Inference \\ signals\end{tabular}} & \multicolumn{3}{c}{Hypotension} & \multirow{2}{*}{SV (mL/min)} 
 & \multirow{2}{*}{SBP (mmHg)} & \multirow{2}{*}{DBP (mmHg)} & \multirow{2}{*}{Age (years)} \\ 
 &  &  & 5-min & 10-min & 15-min &  &  &  &  & \\
\toprule
SL & ECG & ECG & 0.550 & 0.500 & 0.507 & 16.487 (21.013) & 11.625 (14.364) & 8.183 (10.248) & 10.672 (8.505) \\
& PPG & PPG & 0.723 & 0.711 & 0.691 & 16.497 (21.013) & 11.630 (14.364) & 9.034 (10.741) & 10.681 (8.472) \\
& ECG, PPG & ECG, PPG & 0.550 & 0.528 & 0.535 & 16.487 (21.013) & 11.634 (14.364) & 7.722 (9.499) & 10.672 (8.501)  \\ 

Single SSL & ECG & ECG & 0.685 & 0.681 & 0.703 & 16.214 (20.756) & 11.470 (14.210) & 7.689 (9.720)  & 9.292 (7.184) \\
 & PPG & PPG & 0.705 & 0.705 & 0.688 & 16.590 (20.876) & 15.999 (15.439) & 9.566 (11.871) & 12.287 (8.702) \\  

SNUPHY-M & ALL & ECG & 0.716 & 0.688 & 0.711 & 15.946 (20.293) & 11.389 (14.136) & \textbf{7.659 (9.634)} & 9.307 (7.169) \\
& ALL & PPG & 0.806 & 0.771 & 0.749 & 16.063 (20.136) & 11.036 (13.920) & 7.906 (9.961) & 9.862 (7.606) \\
& ALL & ECG, PPG & \textbf{0.832} & \textbf{0.803} & \textbf{0.802} & \textbf{15.487 (19.806)} & \textbf{10.921 (13.134)} & 8.270 (9.027) & \textbf{8.652 (6.399)} \\ \bottomrule
\end{tabular}
}
\label{tbl_full_results_vitaldb}
\end{table*}
Table \ref{tbl_full_results_vitaldb} presents the results of the downstream evaluation performed using the VitalDB open dataset. 
On the VitalDB open dataset, SNUPHY-M outperforms both the SL model and single SSL across all tasks and signal combinations, except for estimating DBP using ECG and PPG signals with the SL model.
With ECG and PPG signals, SNUPHY-M achieves superior AU-ROC of 0.832, 0.803, and 0.802 compared to 0.550, 0.528, and 0.535 of the performance of the SL model for the 5-minute, 10-minute, and 15-minute of hypotension prediction, respectively. 
With ECG signal only, SNUPHY-M also outperforms other models, AU-ROC of 0.716, 0.688, and 0.711, compared to 0.550, 0.500, and 0.507 of SL models, and 0.685, 0.684, and 0.703 of single SSL models. 
Similarly, for the PPG signal only, SNUPHY-M achieves AU-ROC of 0.806, 0.771, and 0.749, exceeding 0.723, 0.711, and 0.691 of the SL models, and also overwhelming 0.708, 0.702, and 0.690 of SSL models (all $p<0.01$).
Furthermore, in the SV prediction task, SNUPHY-M outperforms single SSL and SL models across all signals with the lowest MAE of 15.487 mL/min (all $p<0.01$). 
The SNUPHY-M achieves the MAE of 10.921 mmHg for the SBP estimation and 8.270 mmHg for the DBP estimation, with the SBP estimation being lower than the MAE of the SL model of 11.634 mmHg, while the DBP estimation exceeds the MAE of the SL model of 7.722 mmHg (all $p<0.01$).
Additionally, SNUPHY-M achieves an MAE of 8.652 years for age prediction, significantly better than the SL model, 10.672 years (all $p<0.01$).
In addition, SNUPHY-M consistently outperforms single SSL models across the SV prediction, SBP estimation, and age prediction tasks, further emphasizing its robust and reliable performance.

Table \ref{tbl_full_results_mimic} presents the results of SBP and DBP estimation tasks performed using the MIMIC-III waveform dataset. 
When both ECG and PPG signals are available, SNUPHY-M achieves MAE values of 11.077 mmHg for SBP and 7.689 mmHg for DBP, outperforming the SL model, which shows 11.799 mmHg and 8.264 mmHg, respectively (all $p<0.01$). 
Even when only PPG signals are used, SNUPHY-M demonstrates competitive or superior performance, achieving an MAE of 11.829 mmHg for SBP, compared to 12.363 mmHg for SL and 11.864 mmHg for single SSL (all $p<0.01$). 
Similarly, in estimating DBP using only the input of PPG, SNUPHY-M outperforms both SL and single SSL, achieving an MAE of 7.275 mmHg compared to that of 8.402 mmHg and 7.569 mmHg, respectively (all $p<0.01$).

\begin{table}[]
\caption{Performance comparison for linear probing. MAE for DBP and SBP estimation tasks on the MIMIC-III dataset.}
\centering
\begin{tabular}{@{\hspace{2.5pt}}c@{\hspace{3.5pt}}c@{\hspace{3.5pt}}c@{\hspace{3.5pt}}c@{\hspace{3.5pt}}c}
\toprule
Model         & \begin{tabular}[c]{@{}c@{}}Training \\ signals\end{tabular} & \begin{tabular}[c]{@{}c@{}}Inference \\ signals\end{tabular} & \begin{tabular}[c]{@{}c@{}}SBP \\ (mmHg)\end{tabular} & \begin{tabular}[c]{@{}c@{}}DBP  \\ (mmHg)\end{tabular} \\ \midrule
SL            & ECG & ECG & 11.644 (14.488) & 8.145 (10.733) \\
       & PPG & PPG & 12.363 (14.489) & 8.402 (10.734) \\
       & ECG, PPG & ECG, PPG & 11.799 (14.488) & 8.264 (10.733) \\ 
Single SSL    & ECG & ECG & 11.613 (14.501) & 8.631 (10.893) \\
   & PPG & PPG & 11.864 (14.335) & 7.569 (9.533) \\ 
SNUPHY-M    & ALL & ECG & 12.326 (15.213)  & 7.836 (9.935) \\
  & ALL & PPG & 11.829 (14.547)  & \textbf{7.275 (9.209)}  \\
  & ALL & ECG, PPG & \textbf{11.077 (13.820)} & 7.689 (10.142) \\ \bottomrule
\end{tabular}
\label{tbl_full_results_mimic}
\end{table}

These results indicate that the SNUPHY-M effectively captures complex interrelationships among physiological signals, and encodes both shared and signal-specific features. 
This enables more enriched representations of cardiac function, potentially supporting precise cardiovascular monitoring even when invasive signals like ABP are unavailable.

\begin{table*}[h!]
\caption{Performance comparison of models with subsampling analysis on VitalDB dataset. AU-ROC is used for the hypotension prediction task, and MAE is for the SV prediction task. The test data remains unchanged across all experiments.}
\resizebox{\textwidth}{!}{%
\centering
\begin{tabular}{@{}ccccccccc@{}}
\toprule
\multirow{2}{*}{Model} & \multirow{2}{*}{\begin{tabular}[c]{@{}c@{}}Training \\ signals\end{tabular}} & \multirow{2}{*}{\begin{tabular}[c]{@{}c@{}}Inference \\ signals\end{tabular}} & \multicolumn{3}{c}{\begin{tabular}[c]{@{}c@{}}Hypotension \\ (10-min)\end{tabular}} & \multicolumn{3}{c}{SV (mL/min)} \\
 &  &  & 1\% & 10\% & 100\% & \multicolumn{1}{c}{1\%} & 10\% & 100\% \\ \cmidrule(r){1-9}
SL & ECG & ECG & 0.500 & 0.498 & 0.500 & 49.741 (21.013) & 16.488 (21.013) & 16.494 (21.013) \\
 & PPG & PPG & 0.505 & 0.529 & 0.711 & 50.196 (21.013) & 16.488 (21.013) & 16.493 (21.013) \\
 & ECG, PPG & ECG, PPG & 0.508 & 0.544 & 0.528 & 49.434 (21.013) & 16.557 (21.013) & 16.490 (21.013) \\ 
Single SSL & ECG & ECG & 0.639 & 0.689 & 0.699 & 27.091 (22.931) & 16.591 (21.113) & 16.214 (20.756) \\
 & PPG & PPG & 0.666 & 0.681 & 0.702 & 16.951 (21.471) & 16.754 (21.148) & 16.590 (20.876) \\ 
SNUPHY-M & ALL & ECG & 0.676 & 0.699 & 0.688 & 17.107 (20.962) & 16.076 (20.358) & 15.946 (20.293) \\
 & ALL & PPG & 0.739 & 0.762 & 0.771 & 17.649 (21.381) & 16.309 (20.462) & 16.063 (20.136) \\
 & ALL & ECG, PPG & \textbf{0.781} & \textbf{0.798} & \textbf{0.803} & \textbf{16.502 (20.810)} & \textbf{15.994 (20.244)} & \textbf{15.487 (19.806)} \\ \bottomrule
\end{tabular}
}
\label{tbl_subsampling}
\end{table*}

\paragraph{\textbf{Subsampling analysis}}
In order to assess the efficacy of the SNUPHY-M in scenarios where labeled data for downstream tasks is scarce, experiments were conducted with limited data from the VitalDB open dataset.
As illustrated in Table \ref{tbl_subsampling}, the training and validation data for the downstream hypotension and SV prediction tasks were randomly sampled at a rate of 1\% and 10\%, respectively, while the test data remained consistent.
In comparison to the SL model, which encounters difficulties when confronted with limited ECG and PPG signals, the SNUPHY-M demonstrates remarkable resilience to data scarcity.
In the 10-minute hypotension prediction task using ECG and PPG signals, which is the most commonly monitored and non-invasive signal in clinical practice, the SNUPHY-M achieved an AU-ROC of 0.781 using only 1\% of the downstream data, and 0.798 with 10\%, while the SL model exhibited markedly inferior performance, with an AU-ROC of 0.508 and 0.544, respectively (all $p<0.01$). 
In the SV prediction task, the SNUPHY-M demonstrated the MAE of 16.502 mL/min with 1\% of ECG and PPG signals and 15.994 mL/min with 10\% of ECG and PPG signals, which is substantially lower than the SL model 16.490 mL/min and 15.487 mL/min (all $p<0.01$).
Our findings demonstrate that the representation vector generated by the SNUPHY-M can be employed to construct a resilient predictive model, even in scenarios with a paucity of labeled data.

\begin{table*}[]
\caption{Performance comparison of models with subsampling analysis on the VitalDB dataset. MAE for both tasks. The test data remains unchanged, while only 1\% and 10\% of the total data were used for the training and validation datasets.}
\resizebox{\textwidth}{!}{%
\centering
\begin{tabular}{ccccccccccccc}
\toprule
\multirow{2}{*}{Model} & \multirow{2}{*}{\begin{tabular}[c]{@{}c@{}}Training \\
signals\end{tabular}} & \multirow{2}{*}{\begin{tabular}[c]{@{}c@{}}Inference \\
signals\end{tabular}} & \multicolumn{3}{c}{SBP (mmHg)} & \multicolumn{3}{c}{DBP (mmHg)} & \multicolumn{3}{c}{Age (years)} \\
 &  &  & 1\% & 10\% & 100\% & 1\% & 10\% & 100\% & 1\% & 10\% & 100\% \\
 \midrule
SL & ECG & ECG & \begin{tabular}[c]{@{}c@{}}85.474 (14.364)\end{tabular} & \begin{tabular}[c]{@{}c@{}}11.629 (14.364)\end{tabular} &\begin{tabular}[c]{@{}c@{}}11.625 (14.364)\end{tabular} & \begin{tabular}[c]{@{}c@{}}35.267 (10.248)\end{tabular} & \begin{tabular}[c]{@{}c@{}}8.189 (10.248)\end{tabular} & \begin{tabular}[c]{@{}c@{}}8.183 (10.248)\end{tabular}  & \begin{tabular}[c]{@{}c@{}}20.255 (10.737)\end{tabular} & \begin{tabular}[c]{@{}c@{}}10.673 (8.497)\end{tabular} & \begin{tabular}[c]{@{}c@{}}10.672 (8.505)\end{tabular} \\
 & PPG & PPG & \begin{tabular}[c]{@{}c@{}}85.625 (14.364)\end{tabular} & \begin{tabular}[c]{@{}c@{}}11.628 (14.364)\end{tabular} &  \begin{tabular}[c]{@{}c@{}}11.630 (14.364)\end{tabular} & \begin{tabular}[c]{@{}c@{}}35.650 (10.248)\end{tabular} & \begin{tabular}[c]{@{}c@{}}8.201 (10.248)\end{tabular} & \begin{tabular}[c]{@{}c@{}}9.034 (10.741)\end{tabular}  & \begin{tabular}[c]{@{}c@{}}20.555 (10.823)\end{tabular} & \begin{tabular}[c]{@{}c@{}}10.674 (8.493)\end{tabular} & \begin{tabular}[c]{@{}c@{}}10.681 (8.472)\end{tabular} \\
 & ECG,   PPG & ECG,   PPG & \begin{tabular}[c]{@{}c@{}}85.343 (14.364)\end{tabular} & \begin{tabular}[c]{@{}c@{}}11.631 (14.364)\end{tabular} & \begin{tabular}[c]{@{}c@{}}11.634 (14.364)\end{tabular}  & \begin{tabular}[c]{@{}c@{}}34.659 (10.248)\end{tabular} & \begin{tabular}[c]{@{}c@{}}8.199 (10.248)\end{tabular} & \begin{tabular}[c]{@{}c@{}}7.722 (9.499)\end{tabular}  & \begin{tabular}[c]{@{}c@{}}20.822 (10.903)\end{tabular} & \begin{tabular}[c]{@{}c@{}}10.673 (8.494)\end{tabular} & \begin{tabular}[c]{@{}c@{}}10.672 (8.501)\end{tabular} \\ 
Single   SSL & ECG & ECG & \begin{tabular}[c]{@{}c@{}}46.612 (15.763)\end{tabular} & \begin{tabular}[c]{@{}c@{}}11.794 (14.665)\end{tabular} & \begin{tabular}[c]{@{}c@{}}11.470 (14.210)\end{tabular} & \begin{tabular}[c]{@{}c@{}}10.728 (11.444)\end{tabular} & \begin{tabular}[c]{@{}c@{}}7.819 (9.869)\end{tabular} & \begin{tabular}[c]{@{}c@{}}7.689 (9.72)\end{tabular} & \begin{tabular}[c]{@{}c@{}}11.035 (8.080)\end{tabular} & \begin{tabular}[c]{@{}c@{}}9.342 (7.134)\end{tabular} & \begin{tabular}[c]{@{}c@{}}9.251 (7.165)\end{tabular} \\
 & PPG & PPG & \begin{tabular}[c]{@{}c@{}}22.616 (15.427)\end{tabular} & \begin{tabular}[c]{@{}c@{}}12.049 (14.772)\end{tabular} & \begin{tabular}[c]{@{}c@{}}15.999 (15.439)\end{tabular} & \begin{tabular}[c]{@{}c@{}}10.437 (10.490)\end{tabular} & \begin{tabular}[c]{@{}c@{}}8.661 (10.774)\end{tabular} & \begin{tabular}[c]{@{}c@{}}9.567 (11.871)\end{tabular} & \begin{tabular}[c]{@{}c@{}}10.953 (8.153)\end{tabular} & \begin{tabular}[c]{@{}c@{}}10.664 (8.751)\end{tabular} & \begin{tabular}[c]{@{}c@{}}13.397 (14.317)\end{tabular} \\ 
SNUPHY-M & ALL & ECG & \begin{tabular}[c]{@{}c@{}}14.217 (15.015)\end{tabular} & \begin{tabular}[c]{@{}c@{}}11.490 (14.248)\end{tabular} & \begin{tabular}[c]{@{}c@{}}11.389 (14.136)\end{tabular} & \begin{tabular}[c]{@{}c@{}}8.280 (10.402)\end{tabular} & \begin{tabular}[c]{@{}c@{}}7.745 (9.763)\end{tabular} & \begin{tabular}[c]{@{}c@{}}\textbf{7.659 (9.634)}\end{tabular} & \begin{tabular}[c]{@{}c@{}}10.217 (7.949)\end{tabular} & \begin{tabular}[c]{@{}c@{}}9.424 (7.165)\end{tabular} & \begin{tabular}[c]{@{}c@{}}9.307 (7.169)\end{tabular} \\
 & ALL & PPG & \begin{tabular}[c]{@{}c@{}}17.966 (17.509)\end{tabular} & \begin{tabular}[c]{@{}c@{}}12.356 (15.735)\end{tabular} & \begin{tabular}[c]{@{}c@{}}11.037 (13.920)\end{tabular} & \begin{tabular}[c]{@{}c@{}}9.171 (11.483)\end{tabular} & \begin{tabular}[c]{@{}c@{}}8.307 (10.306)\end{tabular} & \begin{tabular}[c]{@{}c@{}}7.906 (9.610)\end{tabular} & \begin{tabular}[c]{@{}c@{}}10.939 (8.525)\end{tabular} & \begin{tabular}[c]{@{}c@{}}10.975 (8.702)\end{tabular} & \begin{tabular}[c]{@{}c@{}}9.862 (7.606)\end{tabular} \\
& ALL & ECG,   PPG & \begin{tabular}[c]{@{}c@{}} \textbf{11.886 (14.714)}\end{tabular} & \begin{tabular}[c]{@{}c@{}} \textbf{10.784 (13.318)}\end{tabular} & \begin{tabular}[c]{@{}c@{}}\textbf{10.921 (13.134)}\end{tabular} & \begin{tabular}[c]{@{}c@{}}\textbf{7.887 (9.924)}\end{tabular} & \begin{tabular}[c]{@{}c@{}}\textbf{7.421 (9.092)}\end{tabular} & \begin{tabular}[c]{@{}c@{}}8.270 (9.027)\end{tabular} & \begin{tabular}[c]{@{}c@{}}\textbf{9.583 (7.203)}\end{tabular} & \begin{tabular}[c]{@{}c@{}}\textbf{8.728 (6.448)}\end{tabular} & \begin{tabular}[c]{@{}c@{}}\textbf{8.652 (6.399)}\end{tabular} \\
\bottomrule
\end{tabular}
}
\label{appendix_sub_sampling_vitaldb}
\end{table*}

\begin{figure}[t]
\centering
\includegraphics[width=\columnwidth, height=5.8cm]{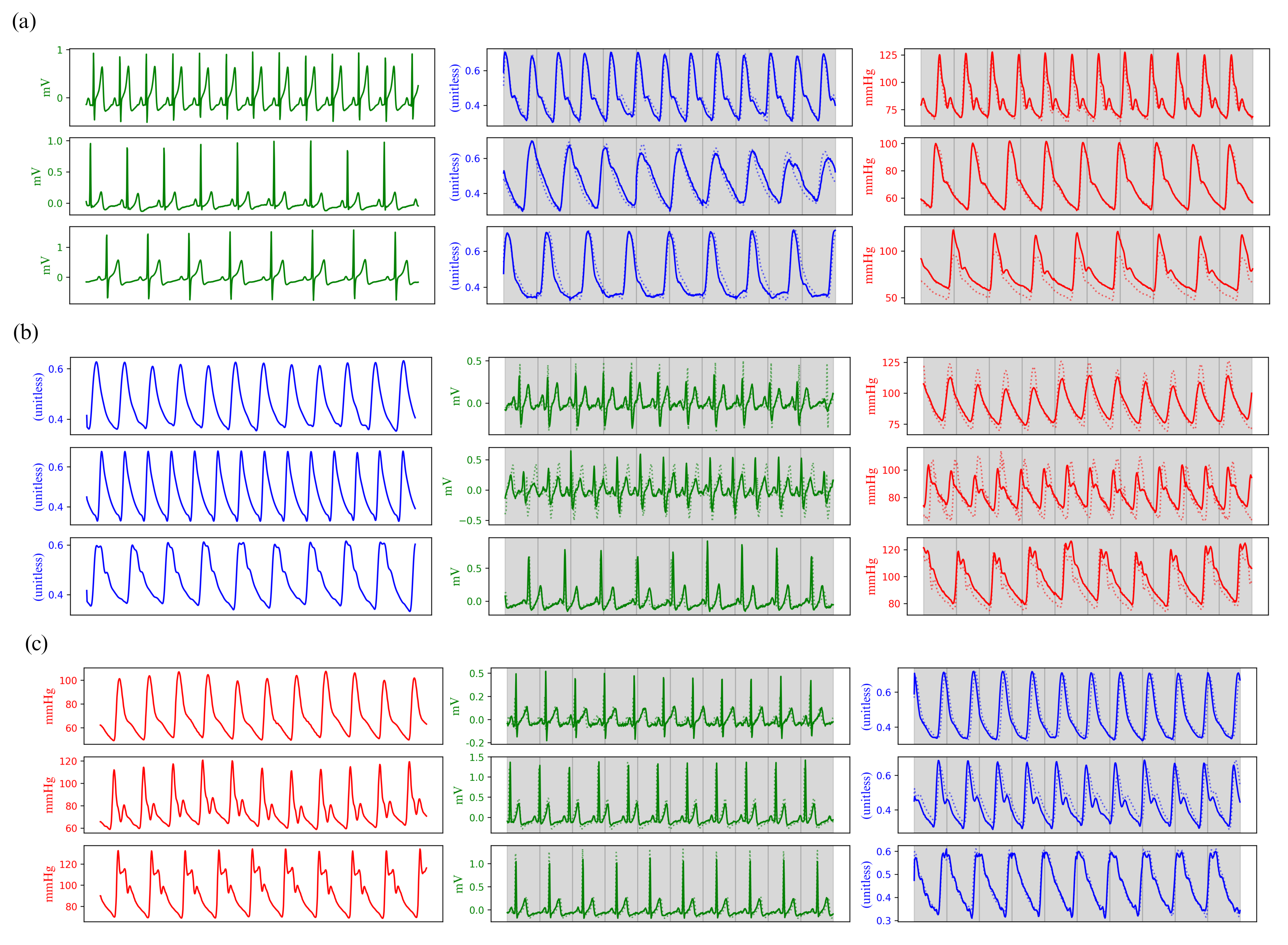}
\caption{SNUPHY-M signal reconstruction across modalities. Input (first column) vs. reconstructed (gray-shaded) signals. ECG=green; PPG=blue; ABP=red.}
\label{fig_reconstruction}
\end{figure}

As shown in Table \ref{appendix_sub_sampling_vitaldb}, the SNUPHY-M consistently outperformed the SL and single SSL models across SBP, DBP estimation, and age prediction tasks, even with limited training data. 
For the SBP estimation task, using ECG and PPG signals, the SNUPHY-M achieved values of 11.886 mmHg and 10.784 mmHg when trained on 1\% and 10\% of the dataset, respectively, compared to the SL model's 85.343 mmHg and 11.631 mmHg. 
Similarly, for the DBP estimation, it recorded values of 7.887 mmHg and 7.421 mmHg under the same conditions, surpassing the SL model's 34.659 mmHg and 8.199 mmHg.
In the age prediction task, the SNUPHY-M yielded MAE values of 9.583 years and 8.728 years, in contrast to the SL model’s 20.822 years and 10.673 years.
These results demonstrate the superior performance and robustness of the SNUPHY-M, even when trained on a small fraction of the available data.

\paragraph{\textbf{Masked raw signals reconstruction}}
Figure \ref{fig_reconstruction} illustrates the reconstruction results of the SNUPHY-M of signal-masking for ECG, PPG, and ABP, respectively.
The model is capable of reconstructing signals by capturing essential physiological characteristics, including the PQRST complex in ECG signals, the distinctive waveform of PPG signals, and the pulsatile nature of ABP signals.
This illustrates the model's ability to comprehend and reproduce the intrinsic physiological processes. 
By employing our model to derive an invasive signal (e.g., ABP) from non-invasive vital signs (e.g., ECG and PPG), we can furnish clinicians with more precise clinical data without unduly distressing the patient.


\subsection{Ablation analysis}
\paragraph{\textbf{Masking strategies}}
Table \ref{tbl_masking_strategy_ablation} presents the experimental results of applying the various masking strategies described in Figure \ref{fig_masking_strategy}.
The proposed masking strategies demonstrate markedly improved predictive performance when all masking strategies are jointly applied, compared to using each masking technique independently or combining only inter- and intra-masking.
The AU-ROC for the 10-minute hypotension prediction improved from 0.860 to 0.879 when signal-masking was added to the combination of inter- and intra-masking, indicating a substantial gain in predictive capability.
Moreover, the MAE for the SBP estimation decreased from 9.275 to 9.092 mmHg, for the DBP estimation from 6.950 to 6.672 mmHg, and for the age prediction from 10.565 to 10.126 years.
This finding indicates that the implementation of signal-masking enhances the model's robustness, leading to a substantial improvement in its predictive proficiency in scenarios with incomplete data.

\paragraph{\textbf{Loss term}}
The effect of the PCC term was assessed in Table \ref{tbl_loss}.
The incorporation of the PCC term significantly enhanced model performance across various tasks, including the hypotension (AU-ROC: 0.780 vs. 0.879, $p<0.01$), the SBP estimation (MAE: 9.351 vs. 9.092 mmHg, $p<0.01$), the DBP estimation (MAE: 6.790 vs. 6.672 mmHg, $p<0.01$) and the age prediction (MAE: 11.067 vs. 10.126 years, $p<0.01$). 
This indicates that understanding the shape or trend of the signal is as crucial as minimizing error in solving various clinical tasks, which is exemplified by the efficacy of the PCC loss term in analyzing physiological signals.

\begin{table}[]
\caption{Ablation analysis of masking strategies (AU-ROC for hypotension; MAE for others; all $p$ < 0.01.).}
\centering
\begin{tabular}{@{}cccccc@{}}
\toprule
Masking strategy & \begin{tabular}[c]{@{}c@{}}Hypotension \\ (10-min)\end{tabular} & \begin{tabular}[c]{@{}c@{}}SBP \\ (mmHg)\end{tabular} & \begin{tabular}[c]{@{}c@{}}DBP \\ (mmHg)\end{tabular} & \begin{tabular}[c]{@{}c@{}}Age \\ (years)\end{tabular} \\ \midrule
 Inter-masking & 0.841 & 9.385 & 7.122 & 11.044 \\ 
 Intra-masking & 0.799 & 9.362 & 7.110 & 10.984 \\ 
 Signal-masking & 0.842 & 9.529 & 7.182 & 10.380 \\ 
 Inter-, intra-masking & 0.860 & 9.275 & 6.950 & 10.565 \\ 
\multicolumn{1}{r}{+ Signal-masking} & \textbf{0.879} & \textbf{9.092} & \textbf{6.672} & \textbf{10.126} \\ 
\bottomrule
\end{tabular}
\label{tbl_masking_strategy_ablation}
\end{table}

\begin{table}[]
\caption{Ablation analysis of PCC in loss function (AU-ROC for hypotension; MAE for others; all $p$ < 0.01.}
\centering
\setlength{\tabcolsep}{4pt} 
\begin{tabular}{@{}ccccc@{}}
\toprule
Loss term & \begin{tabular}[c]{@{}c@{}}Hypotension \\ (10-min)\end{tabular} & \begin{tabular}[c]{@{}c@{}}SBP \\ (mmHg)\end{tabular} & \begin{tabular}[c]{@{}c@{}}DBP \\ (mmHg)\end{tabular} & \begin{tabular}[c]{@{}c@{}}Age \\ (years)\end{tabular} \\ \midrule
RMSE & 0.780 & 9.351 & 6.790 & 11.067 \\
RMSE + PCC & \textbf{0.879} & \textbf{9.092} & \textbf{6.672} & \textbf{10.126} \\ \bottomrule
\end{tabular}
\label{tbl_loss}
\end{table}

\begin{table}[h!]
\caption{Ablation analysis of modules (AU-ROC for hypotension; MAE for others; all $p$ < 0.01.).}
\centering
\begin{tabular}{@{}ccccc@{}} 
\toprule
Method & \begin{tabular}[c]{@{}c@{}}Hypotension \\ (10-min)\end{tabular} & \begin{tabular}[c]{@{}c@{}}SBP \\ (mmHg)\end{tabular} & \begin{tabular}[c]{@{}c@{}}DBP \\ (mmHg)\end{tabular} & \begin{tabular}[c]{@{}c@{}}Age \\ (years)\end{tabular} \\ \midrule
\begin{tabular}[c]{@{}c@{}}w/o signal type \\ embedding\end{tabular} & 0.856 & 9.329 & 6.898 & 10.583 \\
\begin{tabular}[c]{@{}c@{}}w/o cross-attention \end{tabular} & 0.863 & 9.134 & 6.825 & 10.796 \\
\begin{tabular}[c]{@{}c@{}}SNUPHY-M \end{tabular} & \textbf{0.879} & \textbf{9.092} & \textbf{6.672} & \textbf{10.126} \\ 
\bottomrule
\end{tabular}
\label{tbl_modules}
\end{table}

\paragraph{\textbf{Signal type embedding and cross-attention}}
As shown in Table \ref{tbl_modules}, incorporating both signal type embedding and cross-attention contributed significantly to the performance improvements across all tasks. 
Signal type embedding enhanced the model's ability to distinguish and utilize modality-specific features, improving the AU-ROC for hypotension prediction from 0.856 to 0.879 and reducing MAE in SBP, DBP, and age estimation. 
Similarly, cross-attention facilitated the integration of complementary information across signals, leading to further performance gains. 
With its inclusion, the AU-ROC for 10-minute hypotension prediction increased from 0.863 to 0.879, and MAE values decreased consistently across all regression tasks. 
These findings highlight the importance of both components in enabling the model to capture complex inter-signal relationships and extract clinically meaningful representations.

\section{Discussion and conclusion}
In this paper, we proposed the SNUPHY-M that leverages SSL to integrate multiple physiological signals for enhanced cardiac function representation. 
We developed a robust model that outperforms traditional SL and single SSL models using non-invasive physiological signals, particularly when applied to downstream tasks using real clinical data.
The masking strategy is diversified to enhance the representation of complex physiological signals, and the signal-specific decoder is designed to remain functional even when certain signals are missing.
The experimental results demonstrate that the proposed model is effective in scenarios where specific signals are missing or data is limited, by providing representations that incorporate cardiac physiological functions.
It is anticipated that the method will facilitate the development of accurate predictive models in real-world clinical settings with limited data, thereby improving patient outcomes.

\section{Data availability}
Portions of the data and the complete source code used in this study are publicly available to ensure reproducibility. 
This includes all preprocessing scripts, model configuration files, training procedures, and inference code. 
The dataset used for evaluation and detailed instructions for reproducing the experiments can be found at: https://github.com/Vitallab-AI/SNUPHY-M.git.

\bibliography{sample-base.bib}

\begin{thebibliography}{29}
\providecommand{\natexlab}[1]{#1}
\providecommand{\url}[1]{\texttt{#1}}
\expandafter\ifx\csname urlstyle\endcsname\relax
  \providecommand{\doi}[1]{doi: #1}\else
  \providecommand{\doi}{doi: \begingroup \urlstyle{rm}\Url}\fi

\bibitem[Abbaspourazad et~al.(2024)Abbaspourazad, Elachqar, Miller, Emrani, Nallasamy, and Shapiro]{AppleECGPPGCL}
Salar Abbaspourazad, Oussama Elachqar, Andrew Miller, Saba Emrani, Udhyakumar Nallasamy, and Ian Shapiro.
\newblock Large-scale training of foundation models for wearable biosignals.
\newblock In \emph{The Twelfth International Conference on Learning Representations}, 2024.
\newblock URL \url{https://openreview.net/forum?id=pC3WJHf51j}.

\bibitem[Bachmann et~al.(2022)Bachmann, Mizrahi, Atanov, and Zamir]{2022multimae}
Roman Bachmann, David Mizrahi, Andrei Atanov, and Amir Zamir.
\newblock {MultiMAE}: Multi-modal multi-task masked autoencoders.
\newblock In \emph{European Conference on Computer Vision}, 2022.

\bibitem[Bao et~al.(2022)Bao, Dong, Piao, and Wei]{beit}
Hangbo Bao, Li~Dong, Songhao Piao, and Furu Wei.
\newblock {BE}it: {BERT} pre-training of image transformers.
\newblock In \emph{International Conference on Learning Representations}, 2022.
\newblock URL \url{https://openreview.net/forum?id=p-BhZSz59o4}.

\bibitem[Dosovitskiy et~al.(2021)Dosovitskiy, Beyer, Kolesnikov, Weissenborn, Zhai, Unterthiner, Dehghani, Minderer, Heigold, Gelly, Uszkoreit, and Houlsby]{VIT}
Alexey Dosovitskiy, Lucas Beyer, Alexander Kolesnikov, Dirk Weissenborn, Xiaohua Zhai, Thomas Unterthiner, Mostafa Dehghani, Matthias Minderer, Georg Heigold, Sylvain Gelly, Jakob Uszkoreit, and Neil Houlsby.
\newblock An image is worth 16x16 words: Transformers for image recognition at scale.
\newblock In \emph{9th International Conference on Learning Representations, {ICLR} 2021, Virtual Event, Austria, May 3-7, 2021}, 2021.

\bibitem[He et~al.(2022)He, Chen, Xie, Li, Doll{\'a}r, and Girshick]{MAE}
Kaiming He, Xinlei Chen, Saining Xie, Yanghao Li, Piotr Doll{\'a}r, and Ross Girshick.
\newblock Masked autoencoders are scalable vision learners.
\newblock In \emph{Proceedings of the IEEE/CVF conference on computer vision and pattern recognition}, pages 16000--16009, 2022.

\bibitem[Kamanditya et~al.(2024)Kamanditya, Fuadah, Mahardika~T, and Lim]{ecgppgtobp}
Bharindra Kamanditya, Yunendah~Nur Fuadah, Nurul~Qashri Mahardika~T, and Ki~Moo Lim.
\newblock Continuous blood pressure prediction system using conv-lstm network on hybrid latent features of photoplethysmogram (ppg) and electrocardiogram (ecg) signals.
\newblock \emph{Scientific Reports}, 14\penalty0 (1):\penalty0 16450, 2024.

\bibitem[Kouz et~al.(2024)Kouz, Thiele, Michard, and Saugel]{hemodynamiceffect}
Karim Kouz, Robert Thiele, Frederic Michard, and Bernd Saugel.
\newblock Haemodynamic monitoring during noncardiac surgery: past, present, and future.
\newblock \emph{Journal of Clinical Monitoring and Computing}, pages 1--16, 2024.

\bibitem[Kyung et~al.(2023)Kyung, Yang, Choi, Chang, Bae, Choi, and Kim]{attnslppg}
Jehyun Kyung, Joon-Young Yang, Jeong-Hwan Choi, Joon-Hyuk Chang, Sangkon Bae, Jinwoo Choi, and Younho Kim.
\newblock Deep-learning-based blood pressure estimation using multi channel photoplethysmogram and finger pressure with attention mechanism.
\newblock \emph{Scientific reports}, 13, 2023.

\bibitem[Lai et~al.(2023)Lai, Tan, Wang, Ji, Guo, Han, Shi, Feng, and Yang]{wearableecgssl}
Jiewei Lai, Huixin Tan, Jinliang Wang, Lei Ji, Jun Guo, Baoshi Han, Yajun Shi, Qianjin Feng, and Wei Yang.
\newblock Practical intelligent diagnostic algorithm for wearable 12-lead ecg via self-supervised learning on large-scale dataset.
\newblock \emph{Nature Communications}, 14:\penalty0 3741, 2023.

\bibitem[Lee and Jung(2018)]{vitalrecorder}
Hyung-Chul Lee and Chul-Woo Jung.
\newblock Vital recorder—a free research tool for automatic recording of high-resolution time-synchronised physiological data from multiple anaesthesia devices.
\newblock \emph{Scientific reports}, 8\penalty0 (1):\penalty0 1--8, 2018.

\bibitem[Lee et~al.(2022{\natexlab{a}})Lee, Park, Yoon, Yang, Park, and Jung]{vitaldb}
Hyung-Chul Lee, Yoonsang Park, Soo~Bin Yoon, Seong~Mi Yang, Dongnyeok Park, and Chul-Woo Jung.
\newblock Vitaldb, a high-fidelity multi-parameter vital signs database in surgical patients.
\newblock \emph{Scientific Data}, 9\penalty0 (1):\penalty0 279, 2022{\natexlab{a}}.

\bibitem[Lee et~al.(2022{\natexlab{b}})Lee, Lee, Kim, Le, Hao, Jung, Park, Park, Jun, Lee, and Kim]{multiplefeaturestransfusion}
Seung~Mi Lee, Garam Lee, Tae~Kyong Kim, Trang Le, Jie Hao, Young~Mi Jung, Chan-Wook Park, Joong~Shin Park, Jong~Kwan Jun, Hyung-Chul Lee, and Dokyoon Kim.
\newblock Development and validation of a prediction model for need for massive transfusion during surgery using intraoperative hemodynamic monitoring data.
\newblock \emph{JAMA Network Open}, 5\penalty0 (12):\penalty0 e2246637--e2246637, 2022{\natexlab{b}}.

\bibitem[Lee et~al.(2021)Lee, Lee, Chu, Song, Ahn, Lee, Yang, and Koh]{multiplesignalhypotension}
Solam Lee, Hyung-Chul Lee, Yu~Seong Chu, Seung~Woo Song, Gyo~Jin Ahn, Hunju Lee, Sejung Yang, and Sang~Baek Koh.
\newblock Deep learning models for the prediction of intraoperative hypotension.
\newblock \emph{British Journal of Anaesthesia}, 126\penalty0 (4):\penalty0 808--817, 2021.
\newblock ISSN 0007-0912.
\newblock \doi{https://doi.org/10.1016/j.bja.2020.12.035}.

\bibitem[Moody et~al.(2020)Moody, Moody, Villarroel, Clifford, and Silva]{MIMIC}
Benjamin Moody, George Moody, Mauricio Villarroel, Gari~D. Clifford, and Ikaro Silva.
\newblock Mimic-iii waveform database matched subset.
\newblock PhysioNet, 2020.
\newblock URL \url{https://doi.org/10.13026/c2294b}.

\bibitem[Na et~al.(2024)Na, Park, Tae, and Joo]{STMEM}
Yeongyeon Na, Minje Park, Yunwon Tae, and Sunghoon Joo.
\newblock Guiding masked representation learning to capture spatio-temporal relationship of electrocardiogram.
\newblock In \emph{International Conference on Learning Representations}, 2024.
\newblock URL \url{https://openreview.net/forum?id=WcOohbsF4H}.

\bibitem[Nicholls and Shoemaker(1998)]{hemodynamicdefinition}
Tim Nicholls and William~C Shoemaker.
\newblock Recent advances in hemodynamic monitoring and management of the emergency critically ill patient.
\newblock \emph{Current Opinion in Critical Care}, 4\penalty0 (3):\penalty0 168--176, 1998.

\bibitem[Park et~al.(2022)Park, Lee, Jung, and Yang]{attnslmiccai}
Seong-A Park, Hyung-Chul Lee, Chul-Woo Jung, and Hyun-Lim Yang.
\newblock Attention mechanisms for physiological signal deep learning: which attention should we take?
\newblock In \emph{Medical Image Computing and Computer Assisted Intervention Society}, 2022.

\bibitem[Pereira et~al.(2022)Pereira, Ding, Gadhoumi, Tran, Colorado, Meisel, and Hu]{ecgcnnsl}
Tania Pereira, Cheng Ding, Kais Gadhoumi, Nate Tran, Rene~A. Colorado, Karl Meisel, and Xiao Hu.
\newblock Deep learning approaches for plethysmography signal quality assessment in the presence of atrial fibrillation.
\newblock \emph{Physiological measurement}, 40:\penalty0 12, 2022.

\bibitem[Pinsky et~al.(2022)Pinsky, Cecconi, Chew, Backer, Douglas, Edwards, Hamzaoui, Hernandez, Martin, Monnet, Saugel, Scheeren, Teboul, and Vincent]{hemodynamicmonitoringecgppgabp}
Michael~R. Pinsky, Maurizio Cecconi, Michelle~S. Chew, Daniel~De Backer, Ivor Douglas, Mark Edwards, Olfa Hamzaoui, Glenn Hernandez, Greg Martin, Xavier Monnet, Bernd Saugel, Thomas W.~L. Scheeren, Jean-Louis Teboul, and Jean-Louis Vincent.
\newblock Effective hemodynamic monitoring.
\newblock \emph{Critical Care}, 26:\penalty0 294, 2022.

\bibitem[Rahman et~al.(2021)Rahman, Chang, Dong, Conroy, Natarajan, Kinoshita, Vicario, Frassica, and Xu-Wilson]{hemodynamicml}
Asif Rahman, Yale Chang, Junzi Dong, Bryan Conroy, Annamalai Natarajan, Takahiro Kinoshita, Francesco Vicario, Joseph Frassica, and Minnan Xu-Wilson.
\newblock Early prediction of hemodynamic interventions in the intensive care unit using machine learning.
\newblock \emph{Critical Care}, 25:\penalty0 1--9, 2021.

\bibitem[Ramsay(2020)]{ramsay2020non}
Michael~A Ramsay.
\newblock Non-invasive monitoring is coming the full circle, making our patients safer!
\newblock \emph{Journal of Clinical Monitoring and Computing}, 34\penalty0 (5):\penalty0 869--870, 2020.

\bibitem[Schlesinger et~al.(2020)Schlesinger, Vigderhouse, Eytan, and Moshe]{BPprocessingmethod}
Oded Schlesinger, Nitai Vigderhouse, Danny Eytan, and Yair Moshe.
\newblock Blood pressure estimation from ppg signals using convolutional neural networks and siamese network.
\newblock In \emph{ICASSP 2020 - 2020 IEEE International Conference on Acoustics, Speech and Signal Processing (ICASSP)}, pages 1135--1139, 2020.
\newblock \doi{10.1109/ICASSP40776.2020.9053446}.

\bibitem[Shome et~al.(2024)Shome, Sarkar, and Etemad]{aaaippgtoecg}
Debaditya Shome, Pritam Sarkar, and Ali Etemad.
\newblock Region-disentangled diffusion model for high-fidelity ppg-to-ecg translation.
\newblock In \emph{Proceedings of the Thirty-Eighth AAAI Conference on Artificial Intelligence and Thirty-Sixth Conference on Innovative Applications of Artificial Intelligence and Fourteenth Symposium on Educational Advances in Artificial Intelligence}. AAAI Press, 2024.
\newblock ISBN 978-1-57735-887-9.
\newblock \doi{10.1609/aaai.v38i13.29422}.
\newblock URL \url{https://doi.org/10.1609/aaai.v38i13.29422}.

\bibitem[Sun et~al.(2023)Sun, Bresch, Muehlsteff, Schmitt, Long, Bezemer, Paulussen, Noordergraaf, and Aarts]{sbpecgppg}
Shaoxiong Sun, Erik Bresch, Jens Muehlsteff, Lars Schmitt, Xi~Long, Rick Bezemer, Igor Paulussen, Gerrit~J. Noordergraaf, and Ronald~M. Aarts.
\newblock Systolic blood pressure estimation using ecg and ppg in patients undergoing surgery.
\newblock \emph{Biomedical Signal Processing and Control}, 79:\penalty0 104040, 2023.
\newblock ISSN 1746-8094.
\newblock URL \url{https://www.sciencedirect.com/science/article/pii/S1746809422005171}.

\bibitem[Vaid et~al.(2023)Vaid, Jiang, Sawant, Lerakis, Argulian, Ahuja, Lampert, Charney, Greenspan, Narula, et~al.]{BEITECG}
Akhil Vaid, Joy Jiang, Ashwin Sawant, Stamatios Lerakis, Edgar Argulian, Yuri Ahuja, Joshua Lampert, Alexander Charney, Hayit Greenspan, Jagat Narula, et~al.
\newblock A foundational vision transformer improves diagnostic performance for electrocardiograms.
\newblock \emph{NPJ Digital Medicine}, 6\penalty0 (1):\penalty0 108, 2023.

\bibitem[Vincent et~al.(2011)Vincent, Rhodes, Perel, Martin, Rocca, Vallet, Pinsky, Hofer, Teboul, de~Boode, et~al.]{hemodynamicconsensus}
Jean-Louis Vincent, Andrew Rhodes, Azriel Perel, Greg~S Martin, Giorgio~Della Rocca, Benoit Vallet, Michael~R Pinsky, Christoph~K Hofer, Jean-Louis Teboul, Willem-Pieter de~Boode, et~al.
\newblock Clinical review: Update on hemodynamic monitoring-a consensus of 16.
\newblock \emph{Critical Care}, 15:\penalty0 1--8, 2011.

\bibitem[Yang et~al.(2023)Yang, Westover, and Sun]{biot}
Chaoqi Yang, M~Westover, and Jimeng Sun.
\newblock Biot: Biosignal transformer for cross-data learning in the wild.
\newblock In \emph{Thirty-seventh Conference on Neural Information Processing Systems}, 2023.

\bibitem[Yang et~al.(2021)Yang, Jung, Yang, Kim, Shim, Lee, and Lee]{adultapco}
Hyun-Lim Yang, Chul-Woo Jung, Seong~Mi Yang, Min-Soo Kim, Sungho Shim, Kook~Hyun Lee, and Hyung-Chul Lee.
\newblock Development and validation of an arterial pressure-based cardiac output algorithm using a convolutional neural network: Retrospective study based on prospective registry data.
\newblock \emph{JMIR Med Inform}, 9\penalty0 (8):\penalty0 e24762, Aug 2021.

\bibitem[Zhang et~al.(2023)Zhang, Liu, Shi, Chang, Wang, He, and Huang]{MAEFE}
Huaicheng Zhang, Wenhan Liu, Jiguang Shi, Sheng Chang, Hao Wang, Jin He, and Qijun Huang.
\newblock Maefe: Masked autoencoders family of electrocardiogram for self-supervised pretraining and transfer learning.
\newblock \emph{IEEE Transactions on Instrumentation and Measurement}, 72:\penalty0 1--15, 2023.
\newblock \doi{10.1109/TIM.2022.3228267}.

\end{thebibliography}

\appendix
\clearpage
\section{Details in Experimental Settings}

\subsection{Details of datasets}
\paragraph{\textbf{Pre-training datasets}}
From Jan. 2021 to Sep. 2022, ECG, PPG, and ABP signals were collected from adult patients aged 18 years and older who underwent surgery at Seoul National University Hospital. 
Note that ECG lead II was employed.
The data were divided into $80\%$ for training and $20\%$ for validation according to the surgical procedures.
Of the 6,464 surgical procedures, 1,755,522 samples were utilized for training, while 368,825 samples from 1,385 surgical procedures were employed for validation.
Signals were recorded at 500 Hz and resampled to 100 Hz for subsequent analysis.
The signals were divided into 10-sec samples without the use of a sliding window.

\paragraph{\textbf{Downstream datasets}}
The internal dataset comprises ECG, PPG, and ABP signals from 1,386 surgical patients collected between October 2022 and December 2023. 
The VitalDB open dataset includes 3,458 surgical cases recorded via GE monitors \cite{vitalrecorder}, and the MIMIC-III waveform dataset contains 3,658 patients with ECG II, PPG, and ABP recordings. 
All datasets included adult patients aged 18 years or older, and ECG lead II was used to ensure consistency with the pretraining setting.
Signals were originally sampled at 500 Hz (internal and VitalDB) or 125 Hz (MIMIC-III) and were uniformly resampled to 100 Hz. 
Each signal was segmented into non-overlapping 10-sec samples. 
For all tasks, the data were stratified by patient or surgical procedure and allocated into training (80\%) and validation (20\%) sets.

\begin{itemize}[left=0pt]
    \item\textbf{Hypotension prediction} task was accomplished by utilizing 10-sec of input data to predict the occurrence of hypotensive events within 5-, 10-, and 15-min during surgery. 
    Hypotension was labeled as positive if the mean arterial pressure (MAP) dropped below 65 mmHg for a minimum of one minute. 
    Conversely, the label was negative if the MAP remained above 75 mmHg for more than 20-min. 
    Specifically, for the analysis of 5-min hypotension events, the internal dataset included 870 surgical procedures and 135,038 samples, with a hypotension incidence of 36.788\%. Similarly, the VitalDB open dataset consisted of 2,537 surgical procedures and 413,813 samples, showing a hypotension incidence of 23.295\%.
    For 10-min hypotension events, the internal dataset included 844 surgical procedures and 130,928 samples, with a hypotension incidence of 37.297\%. The VitalDB open dataset contained 2,511 surgical procedures and 407,221 samples, showing a hypotension incidence of 25.185\%.
    Finally, for 15-min hypotension events, the internal dataset comprised 808 surgical procedures and 126,619 samples, with a hypotension incidence of 37.481\%. The VitalDB open dataset included 2,466 surgical procedures and 282,679 samples, showing a hypotension incidence of 23.512\%.
    
    \item\textbf{SV prediction} task employed 10-sec of input data immediately preceding the recorded stroke volume during surgery, to estimate the stroke volume.
    Only cases where the stroke volume values ranged between 20–200 mL/min were considered.
    This task was conducted exclusively using the VitalDB open dataset, comprising 727 surgical procedures and 282,679 samples.
    
    \item\textbf{SBP and DBP estimation} included cases where the difference between consecutive BP measurements was less than 20 mmHg and where the difference between SBP and DBP estimation was less than 70 mmHg \cite{BPprocessingmethod}. For the SBP estimation task, only values within the 90–200 mmHg range were considered, and for the DBP estimation task, only values within the 50–120 mmHg range were included.
    For the SBP and DBP estimation tasks, data were collected from three datasets: internal, VitalDB, and MIMIC-III. Specifically, the internal dataset included 1,276 surgical procedures and 378,142 samples, while the VitalDB open dataset comprised 2,790 surgical procedures and 662,070 samples. The MIMIC-III waveform dataset comprised 2,789 patients and 792,188 samples. 
    
    \item\textbf{Age prediction} involved predicting the actual age of patients aged 18 and older. 
    This task used data from both the internal and VitalDB open datasets. The internal dataset included 1,350 surgical procedures and 450,334 samples, while the VitalDB open dataset comprised 2,850 surgical procedures and 815,753 samples.
\end{itemize}

\begin{figure}[t]
\centering
\includegraphics[width=0.9\columnwidth]{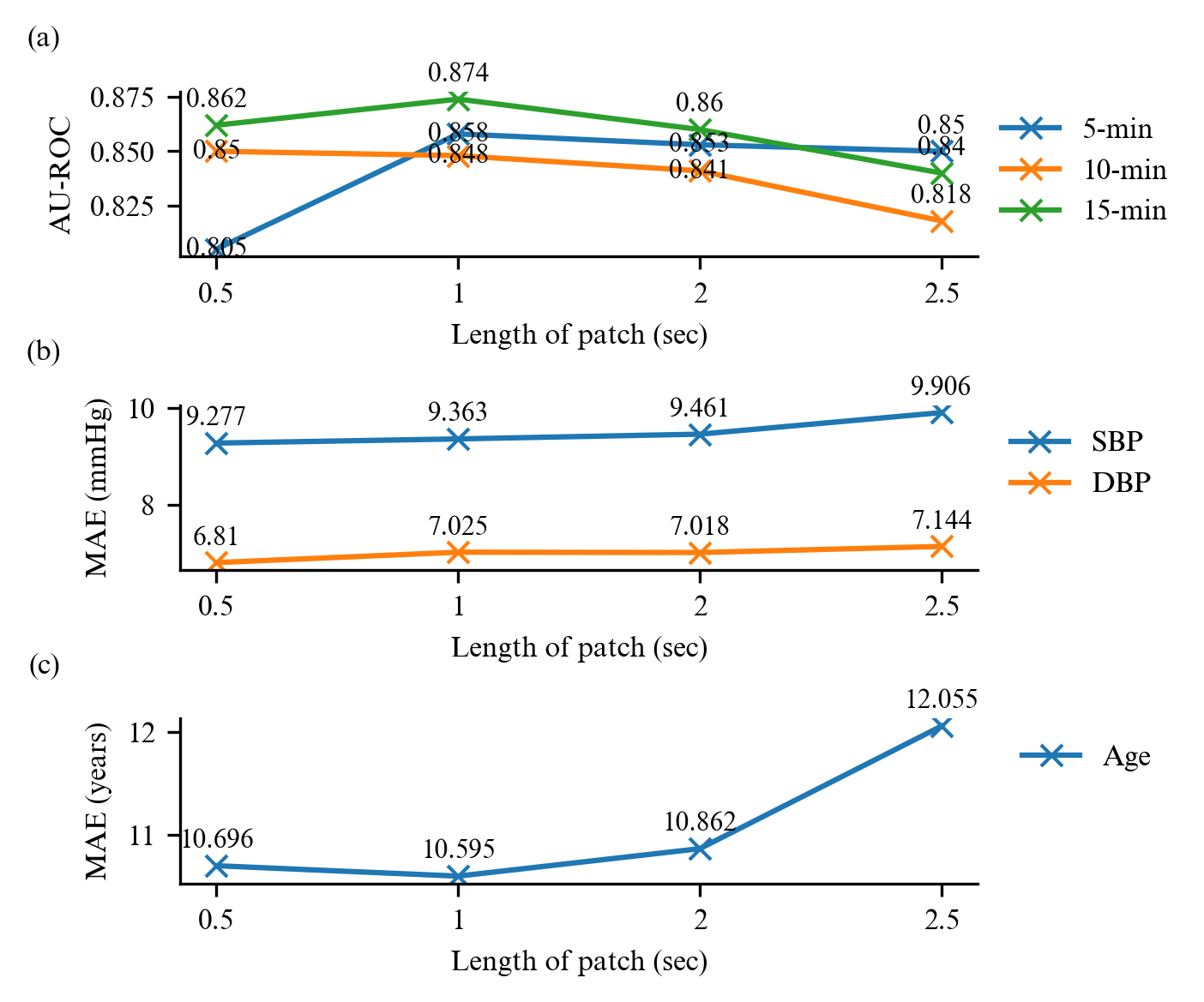}
\caption{Performance evaluation of downstream tasks with varying length of patches. (a) AU-ROC for hypotension prediction across different time windows (5, 10, 15-min). (b) MAE for the SBP and DBP estimation. (c) MAE for the age prediction.}
\label{fig4_segment_size}
\end{figure}

\begin{figure}[t]
\centering
\includegraphics[width=0.9\columnwidth]{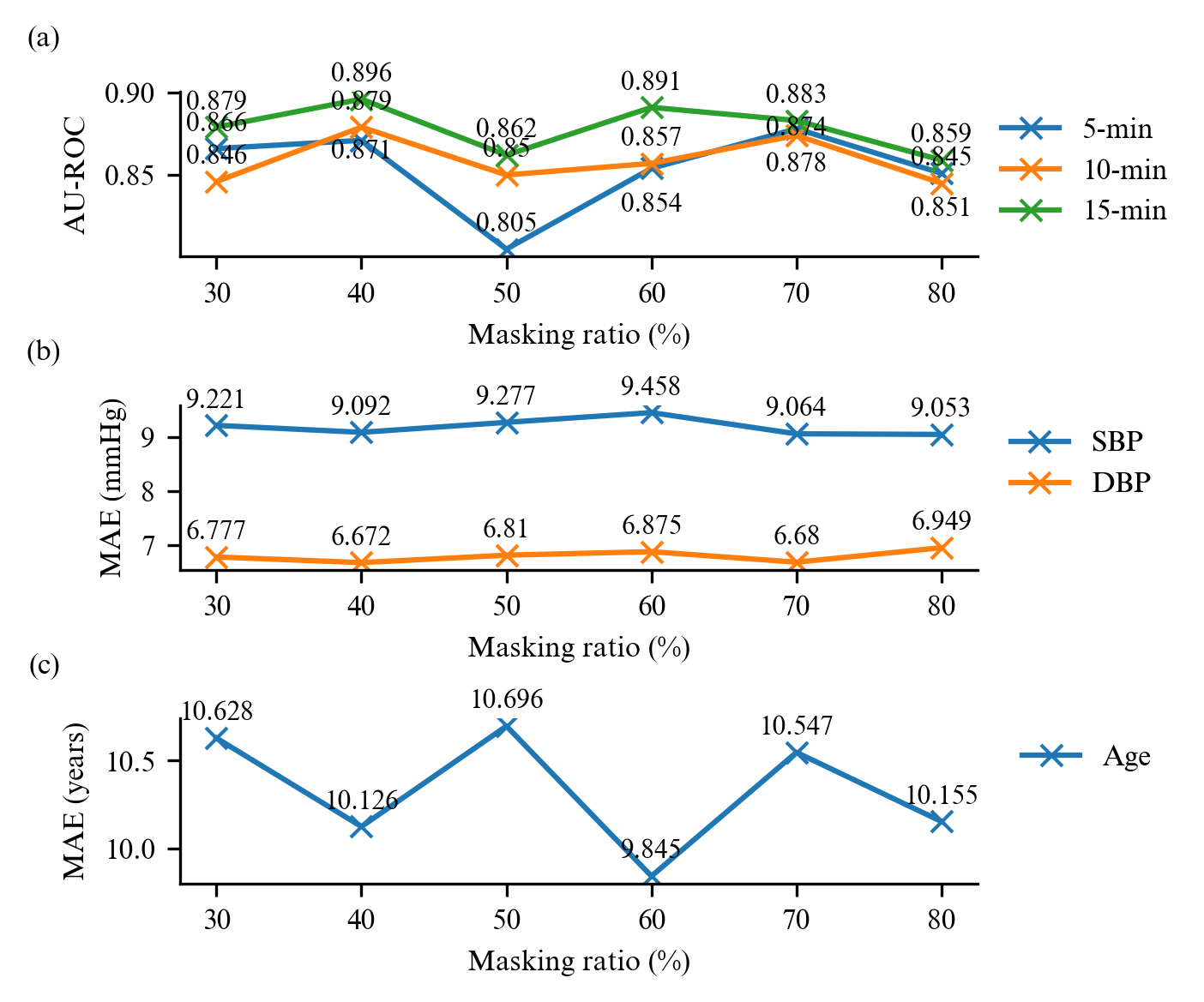}
\caption{Performance evaluation of downstream tasks with varying masking ratios. (a) AU-ROC for hypotension prediction across different time windows (5, 10, 15-min). (b) MAE for the SBP and DBP estimation. (c) MAE for the age prediction.}
\label{fig5_masking_ratio}
\end{figure}

\subsection{Data preprocessing}
Noise and artifacts were removed from ECG, PPG, and ABP signals using bandpass Butterworth filters (0.5–40 Hz for ECG, 0.5–8 Hz for PPG) via NeuroKit2. 
Heart rate (HR) was computed from 10-sec signal segments, and only samples with HR between 60–200 bpm and pulse pressure between 20–100 mmHg were retained. 
Samples with HR discrepancies across modalities exceeding 30 bpm were excluded. 
Beats with a correlation coefficient of at least 0.9 to the average beat in more than 50\% of cases were retained.
ECG and ABP were min-max normalized across the training set, while PPG was normalized per sample to account for device variability.

\subsection{Downstream tasks hyperparameters}
The last fully connected layer was trained with a learning rate of 0.001, a batch size of 1,024, and an Adam optimizer.
The binary cross-entropy was employed for the hypotension prediction task, and the mean squared error was utilized for the other regression tasks.

\section{Additional Experiments}

\paragraph{\textbf{Length of patch}}
The performance of the MAE model is significantly influenced by the choice of length of patch, as it directly impacts the granularity of information captured. 
To determine the optimal length of patch, we conducted experiments with masking ratio fixed at 50\% in scenarios where both ECG and PPG signals were present. 
Lengths of patches of 0.5-, 1-, 2-, and 2.5-sec were evaluated.
Figure \ref{fig4_segment_size}, demonstrated that a length of patch of 0.5-sec yielded the best performance across most tasks. 
Specifically, under the 0.5-sec length of patch, the model achieved an AU-ROC of 0.848 for the 10-min hypotension prediction and 0.874 for the 15-min hypotension prediction, indicating superior predictive capability. 
Additionally, for the SBP and DBP estimation, the model exhibited the lowest MAEs of 9.227 mmHg and 6.810 mmHg, respectively. 
Furthermore, the model achieved an impressive MAE of 10.696 years for age prediction.

\paragraph{\textbf{Masking ratio}}

Masked autoencoder-based models are sensitive to the masking ratio, which affects the difficulty of the reconstruction task and the model's generalization. To examine this effect, we varied the masking ratio from 30\% to 80\%. 
As shown in Figure \ref{fig5_masking_ratio}, a 40\% masking ratio consistently yielded optimal performance across multiple tasks. 
At this setting, the model achieved AU-ROC scores of 0.879 and 0.896 for 10- and 15-min hypotension prediction, respectively, and demonstrated strong regression performance with a DBP MAE of 6.672 mmHg and an SBP MAE of 9.092 mmHg.

\end{document}